\documentclass[12pt]{article}

\usepackage{amsfonts}
\usepackage{amssymb}
\usepackage{amsthm}
\usepackage{upref}

\newcommand{\sss}{\setcounter{equation}{0}}
\newtheorem{theorem}{THEOREM}[section]

\newtheorem{corollary}[theorem]{COROLLARY}

\newtheorem{prop}[theorem]{PROPOSITION}

\newtheorem{example}[theorem]{COUNTER-EXAMPLE}

\newfont{\BBFONT}{msbm10 scaled 1200}
\newcommand{\ere}{ {\mathbb R}}

\newcommand{\ese}{{\mathbb S}}

\def\beq{\begin{equation}}
\def\ene{\end{equation}}

\def \ds {\displaystyle}
\def\qed{\ifhmode\unskip\nobreak\fi\ifmmode\ifinner
\else\hskip5pt\fi\fi\hbox{\hskip5pt\vrule width4pt height6pt
depth1.5pt\hskip1pt}}
\def\scat{s(\omega,\omega^{\prime}; \lambda)}
\def\sc0{s_0(\omega,\omega^{\prime}; \lambda)}
\def\ef{f(\nu,\omega;\lambda)}
\begin{document}
\baselineskip=20 pt
\parskip 6 pt

\title{On Inverse Scattering at a Fixed Energy for   Potentials
 with a Regular Behaviour at   Infinity
 \thanks{ Research partially supported by
Universidad Nacional Aut\'onoma de M\'exico under Project
PAPIIT-DGAPA IN 105799,  by CONACYT under Project P42553­F and by
the European Group of Research SPECT.}}
 \author{ Ricardo Weder\thanks{ †Fellow Sistema Nacional de Investigadores.}
\\Instituto de Investigaciones en Matem\'aticas Aplicadas y en Sistemas \\Universidad Nacional Aut\'onoma de
M\'exico \\
Apartado Postal 20-726, M\'exico DF 01000 \\weder@servidor.unam.mx
\and Dimitri Yafaev \\IRMAR, Universit\'e de Rennes I\\
 Campus de Beaulieu, 35042 Rennes, Cedex, France
\\ yafaev@univ-rennes1.fr}
\date{}
\maketitle
\begin{center}
\begin{minipage}{5.75in}
\centerline{{\bf Abstract}}
\bigskip

We study the inverse scattering problem for electric potentials
and magnetic fields in $\ere^d, d\geq 3$, that    are  asymptotic
sums of   homogeneous terms at infinity. The main result is
 that all these terms can be   uniquely reconstructed from the  singularities in the  forward  direction of the
  scattering amplitude at some positive energy.

\end{minipage}
\end{center}
%\newpage

%%%%%%%%%%%%%%%%
\section{Introduction}\sss
 %%%%%%%%%%%%%%%%

 {\bf 1.1  Background.}
There are different formulations of the inverse scattering problem
for the multidimensional Schr\"odinger operator. The most complete
and difficult problem is to find a one-to-one correspondence
between  potentials on $\ere^d$ and the scattering data. For the
solution of this problem see the paper by L. D. Faddeev \cite{fa1}
and also \cite{hn} and \cite{we4}. A review and further references
can be found in \cite{ne} and \cite{no1}. Here the difficult part
is to find necessary and sufficient conditions on the scattering
amplitude $\ef$, $\nu, \omega \in \ese^{d-1}$, $\lambda >0$, such
that the corresponding potential is local.

A simpler problem of uniqueness and reconstruction of the potential from the scattering data has   different
settings. One of them is uniqueness and reconstruction from the high-energy limit of the scattering amplitude
 which
 was
 considered in \cite{fa} and \cite{be} (see also the books \cite{cs,ne} and the papers \cite{ew,ar}). In
  \cite{fa} it
 is used that the first Born approximation is essentially the Fourier
transform of the electric potential. In the time-dependent method
of \cite{ew, ar}  the X-ray or the Radon transforms of the
electric and the magnetic potentials are reconstructed from the
high-energy limit.

A similar problem of uniqueness and reconstruction of the electric
and magnetic potentials from the scattering amplitude at a fixed
positive energy  was considered in \cite{nus} for potentials of
compact support, and in \cite{no, er, is, uh} for potentials
decaying exponentially at infinity. On the contrary, for general
short-range potentials the scattering matrix at a fixed positive
energy does not determine uniquely the potential. Indeed, in
\cite{cs} examples -in three dimensions- are given of non-trivial
radial oscillating potentials with  decay as $|x|^{-3/2}$ at
infinity such that the corresponding
 scattering amplitude is identically zero at some positive energy. Moreover, in dimension two there are examples
 \cite{Gr} of potentials with a regular decay as $|x|^{-2}$ at infinity that have zero scattering amplitude at
 some positive energy. Nevertheless, if two potentials have the same
 scattering amplitude at some positive energy and they coincide outside of some ball they necessary coincide
  everywhere
 \cite{we}.

  Actually, the same problem appears in different settings. Thus, it is proven in
   \cite{is2, we2, gr}
that the scattering matrix at a fixed positive energy uniquely
determines an exponentially decreasing perturbation of a stratified
media. As another example, we mention that
the scattering matrix at a fixed quasi-energy uniquely determines (see \cite{we3})
time-periodic potentials that decay exponentially at spatial
infinity.

\bigskip

 {\bf 1.2 Setting the problem.}
We consider the related (at least in spirit) problem of  the
unique reconstruction of the asymptotics of the electric and
magnetic potentials if only the singularity of the scattering
amplitude $\ef$ on the diagonal $\nu=\omega $ is known at some
fixed  energy $\lambda>0$. This problem was already studied  in
\cite{js1,js2, j1}, where the methods of microlocal analysis were
heavily used. We mention also a related paper \cite{ni} where
using the method of \cite{ew} the asymptotics of the electric
potential is reconstructed if the scattering amplitude is known on
some (perhaps arbitrarily small) interval of positive energies.

 We
show that this problem admits a quite elementary solution. It
consists of two steps and both of them are in some sense well
known. Let us explain them for the reconstruction of the leading
terms of the electric $V$ and magnetic $A$ potentials. The first
is the small angle approximation to the scattering amplitude $\ef$
(see, e.g., \cite{ll}, \S 127) given by the formula \beq \ef
\approx   e^{-\pi i(d-1)/4} k^{(d-1)/2} (2\pi)^{-(d-1)/2}
\int_{\Pi_{\omega}}\, e^{i\ds k \langle y,\omega-\nu \rangle}  R
(y, \omega   ;\lambda; {\bold V})\, dy, \quad k= \lambda^{1/2},
\label{SM1} \ene where ${\bold V}=(V,A)$, $\Pi_{\omega}$ is the
hyperplane orthogonal to $\omega$,
 \beq
 R(y, \omega ;\lambda ; {\bold V}) = (2ik)^{-1} \int_{-\infty}^{\infty}  ( V(y + t\omega )-2k \langle \omega ,
 A(y
 +t\omega )\rangle  ) \, dt
\label{SM2}\ene
and $\nu\rightarrow\omega$. The second is the classical inversion formula for the Radon transform. Actually,
for any $d$ we use always the two dimensional Radon transform.

\bigskip

 {\bf 1.3 Scattering amplitude.}
Let us recall now the definition of the scattering amplitude. If
the electric, $ V(x)$, and the magnetic, $A(x) $, potentials are
real-valued  functions satisfying the   assumption
\beq
 \left| V(x)\right| +  \left| A(x)\right| +  \left|{\rm div}\, A(x)\right|
  \leq C
 (1+|x|)^{-\rho  }
 \label{1.0} \ene
 where $\rho > (d+1)/2$,
 then the corresponding Schr\"odinger equation
\beq
\left(i \nabla + A(x)\right)^2\psi +V(x)\psi=\lambda \psi, \quad \lambda =k^2>0, \label{Sch}
\ene
has a solution with the asymptotics
 $$
\psi(x,\omega,\lambda)= \hbox{exp}\left(i  k \langle\omega, x\rangle\right) +f |x|^{-(d-1)/2}\,
\hbox{exp}\left(i
 k|x|\right) + o( |x|^{-(d-1)/2})
 $$
 as $|x| \rightarrow \infty$. Moreover, such a solution is unique. The coefficient $f(\nu, \omega;
 \lambda)$ depends on the incident direction $\omega$ of the
 incoming plane wave  $ \hbox{exp} (i k\langle
 \omega,x\rangle )$, its energy $\lambda$  and the direction
 $ \nu=x |x|^{-1}$ of observation of the outgoing
 spherical wave $|x|^{-(d-1)/2}\, \hbox{exp} (i k |x| )$. The function $\ef$
 is known as the scattering amplitude. In terms of the scattering amplitude,
 the scattering matrix $S(\lambda)$ is defined by the formula
  \beq
( S(  \lambda) u )(\omega)=u (\omega)+ i e^{i
 \pi(d-3)/4}\lambda^{(d-1)/4} \left(2\pi\right)^{-(d-1)/2}\int_{\ese^{d-1}}
 f(\nu,\omega; \lambda) u(\omega)d\omega.
\label{ScA}\ene
    A discussion of these   results  can be found, for example, in  \cite{yaf2}.  On the contrary, the scattering
     amplitude can be formally defined by equation (\ref{ScA}) via the scattering matrix (see subs.~2.1, for its
      time-dependent definition) for all short-range potentials when $\rho  > 1$ in (\ref{1.0}). Moreover, if
      the assumption
\beq
 \left|\partial^{\alpha}V(x)\right| +  \left|\partial^{\alpha} A(x)\right|
 \leq C_{\alpha} (1+|x|)^{-\rho -|\alpha|} , \quad \rho  > 1,
 \label{1.1} \ene
is valid for all $\alpha$, then, as shown in   \cite{ag},  $\ef$ is a $C^\infty$-function of  $  \nu,\omega
\in \ese^{d-1}$ away from the diagonal $\nu = \omega$.

\bigskip

 {\bf 1.4 Main results.}
 The diagonal singularity of the scattering amplitude contains all information about the behaviour of
 the potentials $V(x)$ and $A(x)$ as $|x|\rightarrow\infty$. This allows one to recover their asymptotics at infinity.

Indeed,  let us consider first the  leading terms of the
asymptotics of $V(x)$ and $A(x)$ when formulae (\ref{SM1}) and
(\ref{SM2}) can be used. Given the leading singularity of the
scattering amplitude on the diagonal, we can recover from
(\ref{SM1}) the function $R(y, \omega ;\lambda) $ as the inverse
Fourier transform  of this singularity in the hyperplane
$\Pi_\omega$. If $A=0$, then, using (\ref{SM2}),  we can
reconstruct      the asymptotics of $V(x)$ by the inversion
formula for the Radon transform. In the general case, we  take
into account that the integrals
 \beq
R_{e}(y, \omega; V)=  \int_{-\infty}^{\infty} V(y + t\omega )  \, dt, \quad  \omega \in \ese^{d-1}, \quad  y
 \in \Pi_{\omega},
\label{R3}\ene
and
 \beq
R_{m}(y, \omega ; A)=  \int_{-\infty}^{\infty}    \langle \omega ,A(y + t\omega )\rangle   \, dt, \quad
 \omega \in \ese^{d-1}, \quad  y \in \Pi_{\omega},
\label{R5}\ene
  are, respectively,  even and odd functions  of $ \omega \in \ese^{d-1}$,  which determine both  these integrals
  separately.
     This allows us to reconstruct, as before, the electric potential $V$, but it is of course impossible to find the
     magnetic potential $A$ from the integral (\ref{R5}). Nevertheless, we show that this integral determines
     the magnetic field $F={\rm curl}\, A$. One cannot of course   do better because the scattering matrix, as well as
     the integral (\ref{R5}), are invariant with respect to gauge transformations (see subs.~2.2).

Assuming that the potentials are asymptotically homogeneous, we
actually use, for all $d\geq 3$, the inversion of the Radon
transform in  two-dimensional planes. This is important for the
reconstruction of the magnetic field. If $d=2$, then the integral
(\ref{R3}) can be equal to zero for all $y\neq 0$ for a
non-trivial  homogeneous function $V$ (see Counter-example~4.4).
Therefore, in dimension $2$ one cannot hope   to recover the
asymptotics of a potential from the forward singularity of the
scattering amplitude.

Our approach extends to the complete asympotic expansion.
Assume that an electric potential
\beq
V (x) \simeq \sum_{j=1}^\infty V_j(x)
 \label{V}\ene
and a magnetic field
\beq
F (x)\simeq \sum_{j=1}^\infty F_j(x)
 \label{F}\ene
 are asymptotic sums of  homogeneous functions (see subs.~2.3, for precise definitions) of orders $-\rho^{(e)}_j$
 and
$-r^{(m)}_j$, respectively, where $1< \rho^{(e)}_1 < \rho^{(e)}_2   <  \cdots$
and  $2< r^{(m)}_1 < r^{(m)}_2   <  \cdots$. Suppose that all singularities in the  forward  direction  of the
 scattering amplitude at some positive energy are known.
We recover all functions $V_j(x) $ and $F_j(x) $ in a recursive way using at every step the inversion of the
two-dimensional  Radon transform. In particular, our results imply that all asymptotic coefficients $V_j(x) $
and $F_j(x) $ are uniquely determined by  these scattering data.

Compared to papers \cite{js1,js2} we do not assume that the
degrees of homogeneity $ \rho^{(e)}_j=1+j$, $ r^{(m)}_j=2+j$. On
the contrary, we recover these numbers from the singularity of the
scattering amplitude. In particular,  the assumption $ \rho^{(e)}
\geq 2$, $ r^{(m)} \geq 3$ imposed in \cite{js1,js2} turns out to
be unnecessary. Another difference of our work is that  we show
that both sets of functions $V_j$ and $F_j$ can be obtained from
the singularities of the scattering amplitude at only one energy
(not at two as required in \cite{js1,js2}).

We proceed from a complete description of the diagonal singularity
of the scattering amplitude. Such explicit formulae were obtained
in the   papers  \cite{Skr} and  \cite{yaf3, yaf4}. These formulae
generalize (\ref{SM2}) but formula (\ref{SM1}) remains the same.
This shows that actually it is natural to regard the scattering
matrix as a pseudo-differential rather than an integral operator.
This point of view was probably  first explicitly adopted in
\cite{BY} where the asymptotics of the eigenvalues of  the
scattering matrix was found.

Under some minor additional a priori assumptions our results combined with those of  \cite{we} show that the
 electric potential (not only its asymptotic expansion at infinity) is uniquely determined by the scattering
 amplitude at some energy
(see Theorem~4.6). This result implies that the uniqueness theorem does not require the super-power decay of
the potential at infinity.
Of course, it does not contradict the examples of \cite{cs} because we exclude oscillations of the potential
 at infinity and  those of \cite{Gr} because we assume that $d > 2$.

The description of the diagonal singularities of the scattering
amplitude is given in Section~3. The (inverse) problem of the
reconstruction of the asymptotic expansions of $V$ and $F$ from
these singularities  is considered in Section~4. The main result
is formulated as Theorem~4.2.

Finally, we note that our approach extends naturally to
 long-range potentials.

The inverse scattering for the Schr\"odinger equation has many important applications in atomic, molecular
and nuclear physics. Also, the inverse scattering problem at a fixed energy for the Schr\"odinger equation is actually
the same as the inverse scattering problem at a fixed frequency for the wave equation with a variable speed. This problem
has many important applications in applied science and in engineering. For the applications of inverse scattering for the
Schr\"odinger equation and the wave equation see \cite{ps}.

%%%%%%%%%%%%%%%%%%%%
\section{ Basic Notions}\sss
%%%%%%%%%%%%%%%%%%%%%%

{\bf 2.1 Scattering theory.} Here we recall some basic definitions
of scattering theory.
 We consider the Schr\"odinger operator
\beq
H=\left(i \nabla + A(x)\right)^2+V(x)
\label{Schr}\ene
with electric $V(x)$ and magnetic $A(x) =(A^{(1)}(x),\ldots,A^{(d)} (x) )$ potentials in the space $L^2(\ere^d)$
 where $d\geq 2$. Under the condition that $V,A$ as well as
the divergence $\mathrm{div} A$ of $A$ are real and bounded functions the Hamiltonian $H$ is well defined and
self-adjoint
on the Sobolev class $\mathsf{H}^2(\ere^d)$. Let us denote by $H_0=-\Delta$ the ``free" Hamiltonian
corresponding to the case
$V=0$ and $A=0$.  Under   assumption
(\ref{1.0})  where $\rho>1$, the positive spectrum of the operator $H$
is absolutely continuous (and its negative spectrum consists of eigenvalues) and the wave operators
\beq W_{\pm} = \hbox{s-}\lim_{t \rightarrow \pm \infty} e^{ itH}
e^{-itH_0} \label{1.2}\ene
exist and enjoy the intertwining property $H W_{\pm}= W_{\pm} H_0$ (see, for example, \cite{kur} or \cite{RS}).
It follows that the
scattering operator
\beq
 {\mathbf S} = W^{\ast}_+ \, W_-
  \label{1.3} \ene
commutes with $H_{0}$ and   is unitary.

 We are now  able to define the scattering matrix.
Let $\ese^{d-1}$ be the unit sphere in $\ere^d $,
  $\ere^+ = (0,\infty)$ and let $L^2\left(\ere^+, L^2(\ese^{d-1})\right)$
  be the $L^2$-space of  functions defined on $\ere^+$ with values   in $L^2(\ese^{d-1})$. Define the unitary
  mapping
$$
{\cal F}: L^2 (\ere^{d})
\rightarrow
L^2\left(\ere^+, L^2(\ese^{d-1})\right)
$$
by the equation
$$
({\cal F}u )(\omega;\lambda)=2^{-1/2} \lambda^{(d-2)/4} \hat{
u}(\lambda^{1/2} \omega), \quad {\rm where}\quad   \hat{ u}(\xi)=
(2\pi)^{-d/2}\int_{\ere^d} e^{-i\langle x,\xi\rangle}  u(x) dx
$$
 is the Fourier transform of $u$ and the spectral parameter $\lambda$   plays the role of the
 energy of a quantum particle. Then $({\cal F}H_{0}u )( \lambda)=\lambda ({\cal F} u )( \lambda)$ and
$$
({\cal F}{\mathbf S} u )( \lambda)=S(\lambda)  ({\cal F} u )( \lambda).
$$
The unitary operator $S(\lambda): L^2(\ese^{d-1})\rightarrow
L^2(\ese^{d-1})$ is known as  the scattering matrix at energy
$\lambda$.  If assumption (\ref{1.1}) is valid for all $\alpha$,
then, as shown in   \cite{ag},  $S(\lambda)$
 is an integral operator with $C^\infty$-kernel  $ s(\nu ,\omega ; \lambda)$ away from the
 diagonal $\nu = \omega$. Moreover, if
$\rho >(d+1)/2$ in (\ref{1.0}), then   $S(\lambda)$  is related to the
scattering amplitude introduced in Section~1 by formula (\ref{ScA}).
For these results see, for example, \cite{yaf2}.

\bigskip

{\bf 2.2 Magnetic fields.}
It follows from (\ref{1.2}) that the operators
  ${\mathbf S}$ and $S(\lambda)$ are invariant under short-range gauge transformations, $ A
  \mapsto A-\nabla \phi$, where $\phi\in C^\infty (\ere^d)$ and $\partial^\alpha\phi(x)=O(|x|^{-\varrho-|\alpha|})$
   for all $\alpha$ and
   some $\varrho >0$ as $|x|\rightarrow \infty$. Therefore it is natural to associate the scattering matrix
   $S(\lambda)$ directly to the magnetic field $F(x)$. However, since the definition of $S(\lambda)$ was given in
    terms of the magnetic potential $A(x)$, we have to discuss here the relation between  magnetic fields and
     potentials.
Below we use freely three-dimensional notation for the curl and the divergence keeping in mind that in the
general case $A$ is a $1$-form and $F$ is a $2$-form.
 By definition, $F(x)= {\rm curl}\, A(x)$, that is in terms of components
 \beq
F^{(ij)}(x)= \partial_{ i}A^{(j)}(x)-\partial_{ j}A^{(i)}(x).
 \label{curl} \ene
On the contrary, the magnetic potential can be reconstructed from the magnetic field $F(x)$ such that
 $ {\rm div}\, F(x)=0$ only up to arbitrary gauge transformations. It is not convenient to work in the standard
 transversal gauge
$\langle x, A_{tr}(x) \rangle =0$ since even for magnetic fields
of compact support the potential $A_{tr}(x)$ decays  only as
$|x|^{-1}$ at infinity.

Therefore we use the procedure suggested in \cite{yaf} of construction of the short-range magnetic potential
for an arbitrary magnetic field satisfying the condition
 \beq
 \left| \partial^{\alpha} F(x)\right|  \leq C_{\alpha}
 (1+|x|)^{- r-|\alpha|}, \quad r > 2, \quad \forall \alpha.
\label{Fsr} \ene
 Let us introduce the auxiliary potentials
\beq A_{reg}^{(i)} (x) = \int_1^{\infty}s \sum_{j=1}^d
F^{(ij)}(sx)x_j\, ds, \quad A_{\infty}^{(i)} (x) =
-\int_0^{\infty}s \sum_{j=1}^d F^{(ij)}(sx)x_j\, ds.
\label{3.2}\ene Note that $A_{\infty}$ is a homogeneous function
of order $-1$, ${\rm curl}\, A_\infty(x)=0$ for $x\neq 0$ and
$A_{tr}=A_{reg}+A_{\infty}$. Next we define the function $U(x)$
for $x\neq 0$ as a curvilinear integral \beq
U(x)=\int_{\Gamma_{x_{0},x}} \langle A_{\infty}(y), dy \rangle
\label{U} \ene taken between some fixed point  $x_{0} \neq 0$ and
a variable point $x$. It is required that $0\not\in
\Gamma_{x_{0},x}$, so that, in view of   the Stokes theorem, for
$d\geq 3$ the function $U(x)$ does not depend on the choice of a
contour $\Gamma_{x_{0},x}$ and ${\rm grad}\, U(x)=A_\infty(x)$.
Let us now  choose an arbitrary function $\eta \in C^\infty
(\ere^d)$ such that $\eta (x)=0$ in a neighbourhood of zero, $\eta
(x)=1$ for, say,
 $|x|\geq 1$ and set
 \beq
A(x) = A_{reg}(x) + (1- \eta (x)) A_{\infty} (x)-U(x){\rm grad}\, \eta (x) .
\label{AA}\ene
Then $   {\rm curl}\, A(x)=F(x)$, $A \in C^\infty (\ere^d)$ if  $F \in C^\infty (\ere^d)$ and
$A(x) = A_{reg}(x)$ for $|x|\geq 2$. Therefore it follows  from definition (\ref{3.2}) that
under assumption  (\ref{Fsr}) $A(x)$ satisfies estimates (\ref{1.1}) with $\rho=r-1$.

Below we always associate  to a magnetic field $F$  the magnetic
potential $A$ by formulae (\ref{3.2}) --  (\ref{AA}) and then
construct the scattering matrix $S(\lambda)$ in terms of the
Schr\"odinger operator (\ref{Schr}). If another short-range
potential $\tilde{A}$ satisfies the relation $   {\rm curl}\,
\tilde{A}(x)=F(x)$, then necessarily the scattering matrices
corresponding to potentials $A$ and $\tilde{A}$ coincide. This
allows us to speak about the scattering matrix $S(\lambda)$
corresponding   to the magnetic field $F$.

\bigskip

{\bf 2.3 Asymptotic expansions.} Let us introduce some notation. We denote by
$\dot{{\cal S}}^{-\rho} =\dot{\cal S}^{-\rho}({\Bbb R}^d)$ the set
of $C^\infty({\mathbb R}^d\setminus\{0\})$-functions $f(x)$ such that
$\partial^\alphaÊf(x)=O(|x|^{-\rho-|\alpha|})$ as $|x|
\rightarrow\infty$ for all $\alpha$. Then, the standard
H\"ormander class is defined as,  $ {\cal S}^{-\rho}({\Bbb R}^d):=
C^\infty({\Bbb R}^d ) \cap \dot{\cal S}^{-\rho}({\Bbb R}^d)$. An
important example of functions from the class  $\dot{{\cal
S}}^{-\rho}$ are homogeneous functions $f \in C^\infty({\Bbb
R}^d\setminus\{0\})$ of order $-\rho $ such that $f(t x)
=t^{-\rho} f(x)$ for all $x\in {\Bbb R}^d, x\neq 0,$ and $t  > 0$.

Let functions $f_{j}\in \dot{{\cal S}}^{-\rho_{j}}$   where $\rho_{j}\rightarrow\infty$ (but the condition
 $\rho_{j}<\rho_{j+1}$ is not required). The  notation
\beq
f (x) \simeq \sum_{j=1}^\infty f_j(x)
 \label{f}\ene
 means that, for any $N$,   the remainder
 \beq
 f   - \sum_{j=1}^N f_j \in\dot{\cal S}^{-\rho}\quad \mathrm{where} \quad \rho=\min_{j\geq N+1} \rho_{j} .
  \label{fr}\ene
  In particular, if the sum (\ref{f}) consists of a finite number $N$ of terms, then the inclusion (\ref{fr})
   should
  be satisfied for all $\rho$.
A function $f \in C^\infty$ is determined by its asymptotic expansion (\ref{f}) up to a term from the Schwarz
 class ${\cal S}={\cal S}^{-\infty}$.

 It follows from formulae  (\ref{curl}) and (\ref{3.2}) --  (\ref{AA}) that the magnetic field $F(x)$ is
    homogeneous of order $-r^{(m)}<-2$ if and only if the magnetic potential  $A (x)$  is     homogeneous of
    order $-\rho^{(m)}=-r^{(m)}+1<-1$.
  Therefore
$A(x)$ is an asymptotic sum
\beq
A (x) \simeq \sum_{j=1}^\infty A_j(x)
 \label{A}\ene
 of   homogeneous functions $A_j(x)$ of orders $-\rho_{j}^{(m)}$ if and only if  the representation (\ref{F})
   holds
 for   $F(x)$ with homogeneous   functions $F_j(x)$ of orders $-r_{j}^{(m)}$ . Here the coefficients $ F_j(x)$
  and
  $A_j(x)$ are related by formulae  (\ref{curl}) and (\ref{3.2}) --  (\ref{AA}); in particular, $\rho_{j}^{(m)}=
  r_{j}^{(m)}-1$.

Adding terms which are equal to zero, we can assume, without loss
of generality, that the numbers $\rho^{(e)}_j$ and  $r^{(m)}_j$ in
 (\ref{V})  and (\ref{F}) are related by the equality $\rho^{(e)}_j=r^{(m)}_j -1$.
 Anyway, by the reconstruction procedure equality of some terms to zero can never be excluded. Set
${\bold V} (x)=(V(x),A(x))$.
Then expansions  (\ref{V})  and (\ref{F}) are equivalent to the expansion
\beq
{\bold V} (x)  \simeq \sum_{j=1}^\infty {\bold V}_j(x)
 \label{bfV}\ene
 in  homogeneous functions ${\bold V}_{j}(x) =(V_j (x),A_j (x))$ of order  $-\rho_{j}$ where  $\rho_{j} =
 \rho_{j}^{(m)}=\rho_{j}^{(e)}$.

%%%%%%%%%%%%%%%%%%%%
\section{ The Direct Problem}\sss
%%%%%%%%%%%%%%%%%%%%%%

% Let us now be more precise about the formulation of the direct problem.

{\bf 3.1 Singularities of the scattering amplitude.}
Following \cite{yaf3, yaf4},  we give
in this section   a complete description of the diagonal singularity of the integral kernel $ s(\nu,\omega ;
 \lambda)$ of the scattering matrix $S(\lambda)$ for a fixed energy $\lambda>0$.
The dimension $d\geq 2$ is arbitrary. Our goal is to construct, for all $N$, an explicit function $s^{(N)}$
 such   that the difference
\begin{equation}
s-s^{(N)}\in C^{p(N)} (L^2(\ese^{d-1})\times L^2(\ese^{d-1})) \quad {\rm where}\quad
\lim_{N\rightarrow \infty} p(N)= \infty.
 \label{2.15b}\end{equation}
Since $s$ is a $C^\infty$-function  away from the diagonal $\nu =
\omega$, it suffices to construct $s^{(N)}$ in a neighbourhood of
the diagonal only. This construction is given in terms of special
approximate solutions of the Schr\"odinger equation  (\ref{Sch}).
Let us set
 \beq
  u_{\pm}^{(N)}
(x,\xi)=e^{i\langle x,\xi\rangle +i \Phi_{\pm}(x,\hat{\xi})}
{\bold b}_{\pm}^{(N)} (x,\xi),\quad \xi=|\xi |\hat{\xi}\in{\ere}^d,
\label{2.1}\ene
where the phase
\beq
\Phi_{\pm}(x,\hat{\xi}) = \mp \int_0^{\infty} \langle
\hat{\xi},A(x\pm t\hat{\xi})\rangle \, dt
 \label{2.3}\ene
   and
\begin{equation}
   {\bold b}_{\pm}^{(N)}(x,\xi)=\sum_{n=0}^N (2i|\xi|)^{-n} b_\pm ^{(n)}(x,\hat{\xi}).
 \label{2.6}\end{equation}
 The coefficients $ b_\pm ^{(n)}$ are defined here by the recurrent formulae $ b_\pm ^{(0)}(x, \hat{\xi})=1$ and
 \beq
    b_\pm ^{(n+1)}(x,\hat{\xi})  = \mp \int_{0}^{\infty}  f_\pm ^{(n)}(x\pm t \hat{\xi},
   \hat{\xi})\,dt,
   \label{2.9}\ene
   where
\begin{eqnarray}
 f_\pm ^{(n)} & = & 2i \langle A-\nabla \Phi_{\pm}, \nabla
 b_\pm ^{(n)}\rangle-  \Delta  b_\pm ^{(n)}
 \nonumber\\
& +& \left(    \left| \nabla \Phi_{\pm} \right|^2
 - 2\langle A,\nabla \Phi_{\pm}\rangle + V+ |A|^2 + i {\rm div} A -i\Delta \Phi_{\pm}\right)
 b_\pm ^{(n)}.
\label{2.10}
\end{eqnarray}

 One can arrive to these expressions for the coefficients $ b_\pm ^{(n)}$    by plugging
(\ref{2.1}) and(\ref{2.3})  into the Schr\"odinger equation
 (\ref{Sch}) where $\lambda=|\xi|^2$. This yields the  transport equation for the function ${\bold b}_{\pm}^{(N)}
  (x,\xi)$. Seeking it in the form    (\ref{2.6})  and equating coefficients at
the same powers of $(2i|\xi|)^{-n} $, we obtain recurrent
equations for the functions $ b_\pm ^{(n)}$. They are solved by
formulae (\ref{2.9}) and(\ref{2.10}). This is a standard
construction of approximate solutions of the Schr\"odinger
equation going back in the  context of the present paper to
\cite{bu}.
 It is important that the remainder arising when function (\ref{2.1}) is plugged into the Schr\"odinger equation
 satisfies, for larger $N$,  better and better estimates as $|x|\rightarrow\infty$ everywhere except the forward
 direction $\hat{x}=\hat{\xi}$ for the function $u_{+}^{(N)}(x,\xi)$ and except the back direction
 $\hat{x}=- \hat{\xi}$ for the function $u_{-}^{(N)} (x,\xi)$. At the same time the estimates of the remainder
  improve as the energy  $\lambda\rightarrow\infty$.

 A  complete description of the diagonal singularity of the function $s(\nu,\omega ;\lambda)$ is given in the
 following theorem obtained in
 \cite{yaf3, yaf4}.

\begin{theorem}
Let estimates $(\ref{1.1})$   hold for all $\alpha$.
Let $\omega_0\in \ese ^{d-1}$ be an arbitrary point,
$\Pi_{\omega_0}$ be the plane orthogonal to $\omega_0$ and $\Omega=\Omega
(\omega_0,\delta)\subset {\ese}^{n-1}$ be determined by the
condition $ \langle\omega ,\omega_0\rangle>\delta>0$.
  Set $x=\omega_0 z+y,\quad y\in \Pi_{\omega_0}$,
\beq
h_{\pm}^{(N)}(x,\xi)= e^{i\Phi_{\pm}(x,\xi)}\, {\mathbf b}_{\pm}^{(N)}(x,\xi),
\label{2.14x}\ene
\begin{eqnarray}
{\mathbf a}^{(N)}(y, \nu,\omega; \lambda)& =&
 2^{-1}  \langle\nu +\omega ,\omega_0\rangle
\overline{h_+(y, k\nu)} h_-(y,k\omega )
\nonumber\\
& +& (2 i k)^{-1} \Bigl[  \,  \overline{h_+(y, k \nu )} (\partial_z h_-)(y,
k \omega) -
  \overline{(\partial_z h_+)(y, k\nu)} h_-(y, k\omega )
 \nonumber\\
&-&2 i \langle A(y),\omega_0\rangle \overline{h_+(y,
k\nu)}\, h_-(y, k \omega )\Bigr],
 \quad h_{\pm} =h_{\pm}^{(N)}, \quad  k=\lambda^{1/2},
\label{2.15}
\end{eqnarray}
and define
 \beq
 s^{(N)}(\nu,\omega ;\lambda)=
 (2\pi)^{-d+1}  k^{d-1 }
\int_{\Pi_{\omega_0}}e^{ik\langle y
,\omega -\nu \rangle}\, {\mathbf a}^{(N)}(y,
\nu,\omega ; \lambda) \, dy.
\label{2.14}\ene
Then relation $(\ref{2.15b})$ holds.
\end{theorem}

It can be also shown that the $C^{p(N)}$-norm of the kernel
(\ref{2.15b}) is $O(\lambda^{-q(N)})$ as
$\lambda\rightarrow\infty$ where $q(N)\rightarrow\infty$ for $N
\rightarrow\infty$.  Thus, all singularities of $ s(\nu,\omega
;\lambda)$ both for high energies and in smoothness are described
by the explicit formulae (\ref{2.15}), (\ref{2.14}).

Note that the amplitude (\ref{2.15}) is invariant under a gauge
transformation $ A \mapsto  A-\nabla \phi$,  $h_{\pm}
 \mapsto e^{-i\phi}\, h_{\pm}$, where $\phi\in C^\infty$ is an arbitrary function.

  Formula (\ref{2.14}) gives the
  singular part $ s^{(N)}(\nu,\omega ;\lambda)$ of the kernel $s(\nu ,\omega ;\lambda)$  for $\nu,
\omega \in\Omega (\omega_0,\delta)$. Since $\omega_0\in {\ese}^{d-1}$ is arbitrary,
this determines the function   $ s^{(N)}(\nu,\omega ;\lambda)$  for all $\nu,
\omega \in {\ese}^{d-1}$.

\bigskip

{\bf 3.2 Symbol of the scattering matrix.}
 According to (\ref{2.14}) it is natural to regard  the scattering matrix $S(\lambda)$
as a pseudodifferential operator (on the unit sphere) determined
by its  amplitude. For our purposes it is convenient to
reformulate Theorem~3.1 in terms of asymptotic series. First,
replace $ e^{i\Phi_{\pm}(x,\xi)}$ in (\ref{2.14x}) by   its Taylor
expansion. Then, rearrange the representation (\ref{2.15})
collecting together terms of the same power with respect to
${\bold V}=(V,A)$. Thus, the amplitude ${\bold a} (y,\nu, \omega;
\lambda)$ of the pseudodifferential operator $S(\lambda)$ admits
the expansion into the asymptotic series
\[
{\bold a} (y,\nu, \omega; \lambda)  \simeq \sum_{n=0}^\infty {\bold a}_{n} (y,\nu, \omega; \lambda) ,
  \]
  where ${\bold a}_{0}(\nu, \omega; \lambda)= 2^{-1}\langle\nu+ \omega,\omega_{0} \rangle$,  ${\bold a}_{1} $
  depends linearly on ${\bold V}$,  ${\bold a}_{2} $ depends quadratically on ${\bold V}$, etc.  It follows from
   formulae (\ref{2.9}), (\ref{2.10}) that   $  {\bold a}_n \in {\cal S}^{-(\rho -1) n}$ for all $n$ and, moreover,
 \begin{eqnarray}
 2ik {\bold a}_{1}(y, \nu, \omega; \lambda) & =& 2^{-1}\langle \nu+ \omega,\omega_{0} \rangle
 \int_{0}^{\infty}  \Big( V(y + t\nu )+V(y - t\omega )
\nonumber\\
& -& 2k \langle \nu ,A(y +t\nu )\rangle
 -2k \langle \omega ,A(y -t\omega )\rangle   \Big) \, dt,
   \label{2.93ampl}\end{eqnarray}
  up to a term from the class    $   {\cal S}^{-\rho }$.

Applications to the inverse problem require that the pseudodifferential operator $S(\lambda)$ be determined
by its symbol. Let us recall briefly the standard procedure (see, e.g., \cite{ Sh, tr1}) of the passage in
 local coordinates from the amplitude to the corresponding (right) symbol. Let us consider the orthogonal
 projection $p$ of   $\Omega$ on the hyperplane $\Pi_{\omega }$ and set
\beq
 a (y, \omega ; \lambda) \simeq \sum_\alpha  \alpha!^{-1}
   (i k )^{-| \alpha|} \partial^{\alpha}_y\,  \partial^{\alpha}_{\eta}\,
  {\mathbf a} (y, t(\eta), \omega ;\lambda)\Big|_{\eta=p(\omega)}, \quad y\in \Pi_{ \omega},
  \label{2.17}\ene
where the inverse mapping $t=p^{-1}: p(\Omega) \rightarrow \Omega$ is given by the formula $t(\eta)=
(\eta,(1-|\eta|^2)^{1/2})$.    Note that we have taken here
into account the factor $  -k $ in the
phase in (\ref{2.14}) which differs our  definition from the standard terminology. Then
\beq
 s(\nu,\omega;\lambda)= (2\pi)^{-d+1} k^{d-1}
\int_{\Pi_{ \omega}}\, e^{- i  k \langle
y,   \nu  \rangle}  a (y,\omega;\lambda)\, dy
\label{2.18} \ene
so that  the scattering matrix $S(\lambda)$
 can be regarded   as a pseudodifferential operator on $  \ese^{d-1}$ with right symbol $a (y,\omega;\lambda)$.
  This leads to the following conclusion.

\begin{theorem}
Let estimates $(\ref{1.1})$   hold for all $\alpha$.
Then the scattering matrix $S(\lambda)$ is a pseudodifferential operator on the unit sphere. Its right symbol
 $a(y,\omega;\lambda)$ admits the expansion into the asymptotic sum
\beq
 a (y,\omega; \lambda) \simeq 1+ \sum_{n=1}^\infty a_{n}(y, \omega ; \lambda),
  \quad \omega\in\ese^{d-1} ,
\quad y\in\Pi_{ \omega},
  \label{2.17y}\ene
  where   $a_{1} $ depends linearly on ${\bold V} $,  $a_{2} $ depends quadratically on ${\bold V}  $, etc.
 Explicit expressions for all functions $a_{n}$ are obtained by putting
 formulae
 $(\ref{2.15})$ and $(\ref{2.17})$ together. In particular, we have that
   \beq
   a_n \in {\cal S}^{-(\rho -1) n}
   \label{2.92}\ene
 and, moreover,
 \beq
 \tilde{a}_{1}:=  a_1   -   R    \in {\cal S}^{-\rho },
   \label{2.93}\ene
   where the function $ R$ is defined by formula $(\ref{SM2})$. Thus,
    \beq
 a- 1-  R    \in {\cal S}^{-\rho+1-\varepsilon} , \quad   \varepsilon=\min\{\rho-1,1\} ,
   \label{2.94r}\ene
   so that $R(y ,  \omega;\lambda)$ is the principal symbol of the pseudodifferential operator  $S(\lambda)-I$.
 \end{theorem}

 It is important that, for all $n$, the values of $ a_{n} (y,\omega; \lambda)$, $y\in\Pi_{ \omega}$, $y\neq 0$,  are
 determined only by the values of ${\bold V}(x)$ and its derivatives on the line $x=\omega+ty$,   $t\in \ere$, which
 does
  not pass through the origin.

Note that the symbol $a(y,\omega;\lambda)$  can be recovered from
the kernel $ s(\nu,\omega;\lambda)$ by the inversion of the
Fourier transform (\ref{2.18}):
 \beq
a (y,\omega;\lambda)  = \int_{\Pi_{ \omega}}\, e^{ i  k \langle y,
\eta  \rangle}  s( t (\eta),\omega;\lambda) \gamma(\eta)\, d
\eta, \quad y\in\Pi_{ \omega},
\label{2.18in} \ene
where
$\gamma\in C_{0}^\infty (\ere^{d-1})$ is an  arbitrary function
such that $\gamma(\eta)=1$ in some neighbourhood of zero and
$\gamma(\eta)=0$ for, say, $|\eta| \geq 1/2$.

Let  $a$ be the function  constructed in Theorem~3.2.
Let us denote by $T$ the mapping that sends ${\bold V} $ into the function $a-1$. Of course it is defined up
 to a symbol from the class ${\cal S}^{-\infty}$. Thus, we put
 \beq
 T(y,\omega; \lambda; {\bold V})= a(y,\omega; \lambda )-1.
\label{T1} \ene
Moreover, we distinguish the linear part $R$ of $T$ and set
\beq
 Q(y,\omega; \lambda; {\bold V})
 = T(y,\omega; \lambda; {\bold V})-R(y,\omega; \lambda; {\bold V}),
\label{T2} \ene
so that
$$ Q (y,\omega; \lambda; {\bold V})
 =\tilde{a}_{1}(y,\omega; \lambda) + \sum_{n\geq 2}  a_{n}(y,\omega; \lambda).$$

 We need the
   following simple  property of the mapping $Q$.

 \begin{prop}
Suppose that ${\bold V}^{(j)}\in {\cal S}^{-\rho^{(j)}}$, $j=1,2$, where $\rho^{(2)}> \rho^{(1)}> 1$. Then
\[
Q(  {\bold V}^{(1)}+ {\bold V}^{(2)}) - Q(  {\bold V}^{(1)} ) \in {\cal S}^{ -  \rho^{(2)}+1-\varepsilon},
 \quad \varepsilon=\min\{\rho^{(1)}-1,1\}>0.
\]
\end{prop}

Indeed, it follows from (\ref{2.93}) that
\[
\tilde{a}_{1}(  {\bold V}^{(1)}+ {\bold V}^{(2)}) - \tilde{a}_{1}(  {\bold V}^{(1)} )
= \tilde{a}_{1}(  {\bold V}^{(2)} )
 \in {\cal S}^{ -  \rho^{(2)} } .
\]
To consider the  functionals $a_{n}({\bold V})$  for $n\geq 2$,  let us temporarily write them as
$a_{n}({\bold V}) = \mathsf a_{n}({\bold V}, {\bold V},\ldots, {\bold V})$ where $\mathsf a_{n}({\bold V}^{(1)},
 {\bold V}^{(2)},\ldots, {\bold V}^{(n)})$   is linear with respect to its arguments ${\bold V}^{(1)}$,
 ${\bold V}^{(2)}$, \ldots,
$ {\bold V}^{(n)}$. If  $  {\bold V}^{(j)}   \in {\cal S}^{ -  \rho^{(j)}} $ for all $j=1,\ldots, n$, then
$$
 \mathsf a_{n}({\bold V}^{(1)},\ldots, {\bold V}^{(n)})  \in {\cal S}^{ -  \rho^{(1)}-\ldots  -  \rho^{(n)}+n}.
 $$
 This property generalizes (\ref{2.92}). Therefore,
 \[
a_{2}(  {\bold V}^{(1)}+ {\bold V}^{(2)}) - a_{2}(  {\bold V}^{(1)} )
= \mathsf a_{2}({\bold V}^{(1)}, {\bold V}^{(2)}) + \mathsf a_{2}({\bold V}^{(2)}, {\bold V}^{(1)})+
a_{2}(  {\bold V}^{(2)} )
\]
and  the right-hand side belongs to the class $ {\cal S}^{ -  \rho^{(1)}  -  \rho^{(2)}+2} $. The terms
 $a_{n}({\bold V})$ for $n>2$  can be considered quite similarly.

\bigskip

{\bf 3.3  The case of asymptotically homogeneous potentials.}
Let us specify Theorem~3.2 for the perturbations considered in this paper.
 The following result is obtained by plugging  (\ref{V})  and (\ref{A}) into the   expressions
 (\ref{2.15})  and (\ref{2.17}) for the coefficients $a_{n} (y, \omega ; \lambda)$.

\begin{theorem}
Suppose that an electric potential  $V (x)$ and a magnetic field
$F (x)$ are $C^\infty$-functions and that they admit the asymptotic expansions $(\ref{V})$ and
$(\ref{F})$ where $ V_j(x)$ and $ F_j(x)$   are  homogeneous functions of orders $-\rho_j$ and  $-r_j=-\rho_j-1$,
respectively, where $1< \rho_1 < \rho_2   <  \dots$. Let the magnetic potential $A(x)$ be defined
by equalities $(\ref{3.2})$ - $ (\ref{AA})$. Then, the symbol $a  (y,\omega; \lambda)$ of the PDO $S(\lambda)$
 admits
 the expansion into the asymptotic sum
\beq
a  (y,\omega; \lambda)  \simeq 1+ \sum_{ n=1}^\infty \sum_{ m=0}^\infty \sum_{j_{1},j_{2},\ldots, j_{n}}
 a_{n,m; j_{1},j_{2},\ldots, j_{n}}(y,\omega ; \lambda),
  \label{2.17w}\ene
where $j_{k}=1,2,\ldots$ for all $k=1,\ldots, n$, $m=0,1, \dots$, the  functions
$ a_{n,m; j_{1},j_{2},\ldots, j_{n} }(y,\omega ; \lambda)$ only depend on ${\bold V}_{j_{1}}$, ${\bold V}_{j_{2}},
\ldots, {\bold V}_{j_{n}}$
and are   homogeneous of order
$$
n -m-\rho_{j_{1}} -\rho_{j_{2}}-\ldots -\rho_{j_{n}}
$$
in the variable $y$. In particular,
 \beq
   a_{1,0;j}  (y,\omega;\lambda) =R(y,\omega;\lambda; {\bold V}_{j})
      \label{2.94}\ene
      is a homogeneous function of order $-\rho_{j}+1$.
\end{theorem}

\begin{corollary}
If, up to functions from the Schwarz class, $V (x)$ and   $F (x)$ are  homogeneous  $($for large $|x|$$)$
 functions of orders $-\rho$ and  $-r=-\rho -1$, then
\[
a  (y,\omega; \lambda) \simeq 1+ \sum_{ n=1}^\infty \sum_{ m=0}^\infty a_{n,m }(y,\omega ; \lambda),
  \]
where   $a_{n,m }$ is of order $n$ in  ${\bold V} $ $($linear, quadratic, etc.$)$  and is  a  homogeneous
 function  of order
$ n -m  -n\rho $ of the variable $y$.
\end{corollary}

%%%%%%%%%%%%%%%%%%%%%%%%

\section{  The Inverse Problem }
\sss

%%%%%%%%%%%%%%%%%%%%%%%%%

{\bf 4.1 Inversion of the Radon transform.}
Our approach to the  inverse problem relies on the two-dimensional Radon   transform (see, e.g., \cite{hel}).
 Let us recall it briefly here. For $v\in {\cal S}^{-\rho}( \ere^2)$, $\rho>1$, the Radon (or $X$-ray, which is
  the same in the dimension two) transform is defined by the formula
\beq
r(\omega,y;v)= \int_{-\infty}^\infty v(\omega t+y) dt,\quad \omega\in \ese, \quad \langle \omega, y\rangle =0.
\label{R1}\ene
Obviously, $r(\omega,y)=r(-\omega,y)$. The Fourier transform $\hat{v}$  of $v$ and hence the function $v$ itself
 can be recovered in the following way. Let $\omega_{\xi}$ be one of the two unit vectors such that
  $ \langle \omega_{\xi} , \xi\rangle =0$. Then
\beq
\hat{v}(\xi)=(2\pi)^{-1}\int_{-\infty}^\infty e^{- i |\xi | s} r(\omega_{\xi},s\hat{\xi};v) ds,
\quad \hat{\xi}=\xi |\xi |^{-1}.
\label{R2}\ene

We apply this method for the reconstruction of   a homogeneous function $V\in C^\infty (\ere^d\setminus \{0\})$
 of order $-\rho<-1$ from its $X$-ray  transform   (\ref{R3}) known for all $\omega\in \ese^{d-1}$ and
$y\in \Pi_{\omega}$,    $y\neq 0$.
For  an arbitrary $x\in\ere^d \setminus \{0\}$, consider some two-dimensional plane $\Lambda_{x}$ orthogonal to
 $x$ and, for $y\in\Lambda_{x}$, set
$v_{x}(y)=V(x+y)$. Then for   all $\omega\in\Lambda_{x}$, $|\omega|=1$,  and all $y\in\Lambda_{x}$,
 $\langle \omega,y \rangle =0$,
\beq
r(\omega,y;v_{x})= R_{e}(\omega,x+y;V).
\label{R4}\ene
Since $x+y \neq 0$, the function $v_{x} \in {\cal S}^{-\rho}({\ere^2})$ so that we can recover the function $v_{x}$
 and,
in particular, $v_{x}(0)=V(x )$ by formula  (\ref{R2}). This procedure is used  for the reconstruction
of the asymptotics
of the electric potentials.

In the magnetic case we are given only the  integral  (\ref{R5}) for all $\omega\in \ese^{d-1}$ and
$y \in \Pi_{\omega}$,   $y\neq 0$.
Since this integral is zero  if $A(x)={\rm grad}\, \phi(x)$  and
$\phi(x)\rightarrow 0$ as $|x|\rightarrow \infty$, we cannot hope to recover $A$ from this equation.
Nevertheless the corresponding magnetic field $F(x)={\rm curl}\, A(x)$ can be recovered. Assume
 again that $A\in C^\infty (\ere^d\setminus \{0\})$  is a
homogeneous vector-valued function  of order $-\rho<-1$. Let us consider one of the components of $F(x)$,
for example, $F^{(12)}(x)$. We will first show how $F^{(12)}(x)$ can be reconstructed everywhere except
 the plane  $L_{12}$ where $x_{3}=\ldots= x_{d}=0$. Let $\omega =
(\omega_1,\omega_2,0, \ldots, 0)$ be any unit vector in the
plane $L_{12}$,  let $ \nu= (-\omega_2, \omega_1,0,\ldots,0)$ be the unit
vector obtained by rotating $\omega$ in the plane $L_{12}$  by the angle $\pi/2$ in the
counter-clockwise sense and let  $\tilde{x}=(0,0,x_3,\ldots,x_n)\neq 0$  be an arbitrary vector
 that is orthogonal to   $L_{12}$. Set $f_{\tilde{x}}^{(12)} (y)=F^{(12)} (\tilde{x}+y)$  for $y\in L_{12}$.
 It is easy to see that
\beq
r(\omega,y; f_{\tilde{x}}^{(12)})
=- \partial  R_{m}(\omega, s\nu +\tilde{x} ;A)/\partial s,
\quad y=s\nu.
\label{3.6}\ene
Indeed, since $F^{(12)} $ is invariant under   rotations   in the  plane $L_{12}$, it suffices to check
 (\ref{3.6}) for the case $\omega_1=1$, $\omega_2=0$ when (\ref{3.6}) reads as
$$
\int_{-\infty}^{\infty}  F^{(12)} (t, s ,x_3,\ldots,x_n ) dt =
 -\frac{\partial}{\partial s} \int_{-\infty}^{\infty}  A^{(1)} (t, s,x_3,\ldots,x_n )dt.
$$
For the proof of this relation, we have only to use the definition
$F^{(12)}=\partial A^{(2)}/ \partial t - \partial A^{(1)}/ \partial s$ and the fact that the integral
of $\partial A^{(2)}/ \partial t $ is zero. Finding expression (\ref{3.6}), we can recover
$F^{(12)} ( y +\tilde{x})$ for all $y\in L_{12}$ by formula (\ref{R2}). By virtue of the homogeneity
 of the function
$F^{(12)}$, this yields   $F^{(12)}(x)$ everywhere except the
plane  $L_{12}$. Thus, the magnetic field $F (x)$ is reconstructed
for all $x$ such that all coordinates $x_1,x_2 \ldots, x_d$ are not equal to zero. This
condition can be achieved for any arbitrary $x\neq 0$ by a suitable choice of the coordinate system.

 Formula (\ref{3.6}) can   of course be
rewritten in the invariant way. For example, in the case $d=3$ it
means that, for arbitrary $\omega, n\in \ese^2$, $\langle \omega,
n \rangle=0$,
$$
\int_{-\infty}^{\infty}  \langle n, {\rm curl}\, A (t \omega +x )\rangle dt
=
\langle \omega\wedge n, \nabla_{x}\int_{-\infty}^{\infty}  \langle \omega , A (t \omega +x )\rangle dt \rangle,
$$
where the symbol $``\wedge"$ means the vector product.

Finally, if only the combination (\ref{SM2}) is known, then using that $R_{e}$ and $R_{m}$ are even
 and odd functions of $\omega\in \ese^{d-1}$, respectively, we obtain that
 \begin{eqnarray*}
R_{e}(y, \omega ;V ) &=&  ik (R(y, \omega;\lambda;{\bold V}) + R(y,- \omega;\lambda;{\bold V})) ,
\\
 R_{m}(y, \omega ;A)& =&  i 2^{-1}( R(y, -\omega;\lambda;{\bold V}) - R(y, \omega;\lambda;{\bold V})) .
 \end{eqnarray*}
 This allows us to reconstruct  the functions $V $ and $F $ by the formulae given above.

  In particular, we obtain

  \begin{prop}
Let $V$ and $F$ be homogeneous functions on $\ere^d,d\geq 3,$ of
orders $-\rho<-1$ and $-r=-\rho-1$. If $R(y, \omega;\lambda;{\bold
V})=0$ for all $\omega\in\ese^{d-1}$ and $y\in \Pi_{\omega}$, $y
\neq 0$, then $V(x)=F(x)=0$.
\end{prop}

\bigskip

{\bf 4.2  Reconstruction of the potential at infinity.}
Now we are in a position to reconstruct, under the assumptions of Theorem~3.4, all asymptotic coefficients
$V_{j}$ and $F_{j}$ in (\ref{V}) and (\ref{F})  from the kernel $s (\nu ,\omega; \lambda)$ of the scattering
 matrix known, up to a $C^\infty( \ese^{d-1} \times  \ese^{d-1})$-function,  for  some $ \lambda>0$ and all $\nu ,
 \omega\in \ese^{d-1}$.
Thus, we assume that $V$ and $F$ admit the asymptotic expansions (\ref{V}) and (\ref{F}). Then  (\ref{bfV}) holds
 but the functions  ${\bold V}_{j}$, and, in particular,  their orders of homogeneity $-\rho_{j}$, are not known.
 Let us   define the function $a(y,\omega; \lambda)$
 by formula (\ref{2.18in}), that is $a(y,\omega; \lambda)$ is the right symbol of the pseudodifferential operator
 $S(\lambda)$.  This function   admits an asymptotic expansion as in (\ref{2.17y}) where $a_{0}=1$ and $a_{n}$ are
 homogeneous functions of order $-\mu_{k}$ and $0 <\mu_{1} <\mu_{2}<\ldots$.
   Our goal is to solve   equation  (\ref{T1}), where the function $a$ is known with respect to $\bold V$. Of
    course
    both sides of (\ref{T1}) are defined up to   functions from the Schwarz class ${\cal S}$.
     Under the assumptions of Theorem~3.4 the orders of homogeneity in the left-
    and right-hand sides of this equation, as well as the terms of the same order,  should coincide.

   Let us compare terms of the highest order in (\ref{T1}). Taking into account relation (\ref{2.94r}),
    we see that
    $\rho_{1}=\mu_{1}+1$ and
   \beq
   R (y, \omega;\lambda; {\bold V}_{1}) =  a_{1}   (y,\omega;\lambda).
   \label{2.94inv}\ene
 This allows us to reconstruct  the coefficients $V_{1}$ of the electric potential   and $F_{1}$  of the
  magnetic field  by the formulae of the previous subsection. Then we find $A_{1}$ by formulae (\ref{3.2}) - (\ref{AA}).

 Below it is convenient to introduce the following notation. Suppose that some function
  $f$ admits the expansion in the asymptotic series (\ref{f}) of   homogeneous functions. By taking sums of such
   functions it cannot be excluded that some terms will be equal to zero. Therefore we define  $f^\sharp $ as the
   highest order homogeneous term
   $f_{k}$ in  (\ref{f})  that is not identically zero. For example, $f^\sharp =0 $ if $f\in {\cal S}$.

 Suppose now that we have found the coefficients   ${\bold V}_{k}$     for all $k=1, \ldots, n-1$, $n\geq 2$.
 Let us reconstruct ${\bold V}_n$. We apply Proposition~3.3 to the functions
 ${\bold V}^{(1)}=\sum_{j=1}^{n-1}{\bold V}_{j}$, ${\bold V}^{(2)}={\bold V}-{\bold V}^{(1)}$
 where $\rho^{(1)}=\rho_{1}$,  $\rho^{(2)}=\rho_{n}$. This yields
 \[
 Q({\bold V})-  Q(\sum_{j=1}^{n-1}{\bold V}_{j}) \in \dot{\cal S}^{  -\rho_{n}+ 1-\varepsilon}, \quad \varepsilon>0,
 \]
 and  thus,   for an arbitrary $\rho_{n} $, this term can be neglected compared to $R ( {\bold V}_{n})$.
 All terms $R ( {\bold V}_{j})$, $j\geq n+1$, are also negligible  compared to $R ( {\bold V}_{n})$. Therefore
 rewriting equation (\ref{T1}) as
 \[
    R (  {\bold V}_{n}) + \sum_{j\geq n+1}  R (  {\bold V}_{j})
+(    Q (  {\bold V} )- Q ( \sum_{j=1}^{n-1}{\bold V}_{j})) =  a-1   - T(  \sum_{j=1}^{n-1}{\bold V}_{j})
    \]
    and selecting terms of the highest order, we obtain that
 \beq
    R (  {\bold V}_{n})  = ( a-1    - T ( \sum_{j=1}^{n-1}{\bold V}_{j}))^\sharp.
     \label{inv3}\ene
Having found $ R (  {\bold V}_{n}) $,  we can, similarly to (\ref{2.94inv}),
   recover  the functions $V_{n}$, $F_{n}$ and, consequently,
   using formulae (\ref{3.2}) - (\ref{AA}) the corresponding $A_{n}$.

Thus, we have arrived at our main result.

\begin{theorem}
Suppose that an electric potential  $V (x)$ and a magnetic field
$F (x)$ are $C^\infty$-functions on $\ere^d,d\geq 3,$ and that they admit the
asymptotic  expansions $(\ref{V})$ and $(\ref{F})$ where $ V_j(x)$
and $ F_j(x)$   are     homogeneous functions of orders $-\rho_j$
and
 $-r_j=-\rho_j-1$, respectively, where $1< \rho_1 < \rho_2   <  \cdots$. Let the magnetic
  potential $A(x)$ be defined
by equalities $(\ref{3.2})$ - $(\ref{AA})$.
 Then  the scattering data consisting of
the kernel $\scat$ of the scattering matrix  at a fixed positive
energy  in a neighbourhood of the diagonal $\nu=\omega$
uniquely determines each one of the $ V_j(x)$ and the $ F_j(x)$.
Moreover, the functions $ V_1(x)$ and   $ F_1(x)$ can be reconstructed from formula
$(\ref{2.94inv})$ and   the functions $ V_j(x)$ and   $ F_j(x)$ for $j\geq 2$ can be recursively
reconstructed from formula
 $(\ref{inv3})$.
\end{theorem}

\begin{corollary}
If kernel of the operator $S(\lambda)-I$ belongs, for some $\lambda>0$, to the space
$ C^\infty(\ese^{d-1}\times \ese^{d-1})$, then $V\in {\cal S} ({\ere}^{d})$ and  $A\in {\cal S}
({\ere}^{d})$. More generally,
let  $V^{(1)}, F^{(1)}$ and  $V^{(2)}, F^{(2)}$ satisfy the assumptions of Theorem~$4.2$.
Suppose that   $s^{(1)}(\lambda)-s^{(2)}(\lambda)\in C^\infty(\ese^{d-1}\times \ese^{d-1})$
 for some $\lambda > 0$. Then
$V^{(1)}-V^{(2)} $ and $F^{(1)}-F^{(2)}$ belong to the Schwarz class ${\cal S} ({\ere}^{d})$.
\end{corollary}

Suppose that, for some $n\geq 2$,
\[
 a-1    - T ( \sum_{j=1}^{n-1}{\bold V}_{j})\in {\cal S} (\Pi_{\omega})
\]
for all $\omega\in {\ese}^{d-1}$. Then it follows from (\ref{inv3}) that $ V_j=F_{j}=0$ for all $j\geq n$.

 We emphasize that,  unlike $\rho_{1}=\mu_{1}+1$, other orders    $-\rho_{n}$ are not
 determined by the sequence
 $\mu_{1}, \mu_{2}, \ldots$ only, in particular, because a cancellation of
  different terms in the right-hand side of
 (\ref{inv3}) might occur. For example, in
 \beq
    R (  {\bold V}_{2})  = (a -1   - T ( {\bold V}_{1}))^\sharp
    =( a -1  -a_{1}    - Q ( {\bold V}_{1}))^\sharp
     \label{inv2}\ene
a cancellation of terms
  in $a -1  -a_{1} $ and $Q (  {\bold V}_{1})$ cannot be excluded. Nevertheless $-\rho_{2}$
  should be smaller  than the maximal order of these two series. The order
of $a -1  -a_{1} $    equals $-\mu_{2}$ and the order of
  $Q (  {\bold V}_{1})$   equals or is smaller than $-\min\{   \rho_{1}, 2\rho_{1}-2\}$.   This yields the bound
  \[
  \rho_{2}\geq \min\{ \mu_{2}, \rho_{1}, 2\rho_{1}-2\}
  =\min\{ \mu_{2}, \mu_{1}+1, 2\mu_{1} \}.
  \]

\bigskip

{\bf 4.3 A counter-example.}
We give now a counter-example that shows that in the dimension two the
argument above does not work. Let us first calculate the Radon transform of the function
 $V (x) =  v(\hat{x}) |x|^{-2}$ where $\hat{x}= x |x|^{-1}\in \ese $. Making
 the change of variables $t= |y| \tau$,  $\tau= \tan s$, we see that
\begin{eqnarray}
   \int_{-\infty}^{\infty}  V(y + t \omega)\,  dt& =&|y|^{-1}
    \int_{-\infty}^{\infty}  v\left(\frac{\hat{y} + \tau\omega}{\sqrt{\tau^2+1}}\right) \frac{d\tau }{\tau^2+1} =
|y|^{-1}     \int_{-\pi/2}^{\pi/2}   v (\cos s \hat{y}  + \sin s\omega ) ds
\nonumber\\
&=& |y|^{-1} \int_{\ese^\pm ( \omega)} v(\theta) d\theta, \quad \langle \omega, y  \rangle=0, \quad y\neq 0,
\quad \mathrm{if}  \quad \hat{y}\in \ese^\pm ( \omega),
\label{RaCi0}\end{eqnarray}
where $\ese^+( \omega)$ ($\ese^-( \omega)$) is the half-circle passed
 from the point $-\omega$ to $\omega$ in the counter-clockwise (clockwise) direction.

It follows that the Radon transform of $V$ is equal to zero for all $\omega\in \ese$, $\langle \omega, y
 \rangle=0$, $y\neq 0$,
   if and only if
\[
 \int_{\ese^+ ( \omega)} v(\theta) d\theta=0
 \]
 for all $\omega\in \ese $.
 It is easy to see that this condition is equivalent to two conditions
 \beq
v(\omega)=v(-\omega) \quad \mathrm{and} \quad
 \int_{\ese  } v(\theta) d\theta=0.
 \label{RaCi1}\ene
  For example, they are satisfied if
  $$
  v(\omega)=2 \langle \omega,\omega_{0}  \rangle^2-1
  $$
  where $\omega_{0}\in \ese $ is some fixed vector.

\begin{example}{\rm
 Let  $d=2$,  let $A =0$ and  let the electric potential  $V$ be a $C^\infty$-function    such that  $V (x) =
  v(\hat{x}) |x|^{-2}$ for sufficiently large $|x|$.  Assume that the function $v$ satisfies  conditions
  (\ref{RaCi1})  but is not identically zero.
  Then the Radon transform (\ref{RaCi0}) of $v(\hat{x}) |x|^{-2}$   equals zero
   for all $\omega \in \ese $, $\langle \omega,y \rangle=0$, $y\neq 0$,   although $v$ does not. According
   to Theorem~3.2, in this case $S(\lambda)-I$ is the PDO from the class ${\cal S}^{-2}$.}
\end{example}

 We emphasize that in this example the leading singularity of the scattering amplitude disappears
 for all energies $\lambda>0$. The examples of \cite{Gr} are much stronger in the sense that the
  scattering amplitude (not only its leading singularity) equals zero.
On the other hand, this is true for one energy only. It is interesting that both
the examples above and  of  \cite{Gr} are two-dimensional and the potentials decay at infinity as
homogeneous functions of order $-2$.

\bigskip

{\bf 4.4 Uniqueness result.}
In this subsection we assume that the magnetic field is zero and   that $d\geq 3$.
According to Corollary~4.3 the electric potential  is determined by   the scattering matrix
 $S(\lambda)$  for some $\lambda > 0$ up to a term  from the Schwarz class. Let us show now
 that under stronger a priori assumptions this remainder  can be removed.  To that end we need
the following result which is a particular case of Theorem 1 of
\cite{we}.

\begin{theorem}
Let the potentials $V^{(1)}$ and  $V^{(2)}$ on $\ere^d,d\geq 3,$ satisfy estimates
$(\ref{1.0})$ with $\rho  > d$ and  let $V^{(1)}(x)=V ^{(2)}(x)$ for sufficiently large $|x |$.
Suppose that the corresponding scattering matrices $S^{(1)}(\lambda)=S^{(2)}(\lambda)$  for some $\lambda > 0$. Then
$V^{(1)}(x)=V^{(2)}(x)$ for all $x\in\ere^d$.
\end{theorem}

Combining Theorems~4.2 and 4.5,  we obtain a new uniqueness result.

\begin{theorem}
Let $C^\infty$-potentials $V^{(k)}$, on $\ere^d,d\geq 3, k=1,2$,   equal $($finite$)$ sums of homogeneous functions of
 order  $-\rho_{j}^{(k)}$ for sufficiently large $|x|$.  Let $\rho_{j}^{(k)}>d$ for $k=1,2$ and all $j$.
Suppose that the corresponding scattering matrices $S^{(1)}(\lambda)=S^{(2)}(\lambda)$  for some $\lambda > 0$. Then
$V^{(1)}(x)=V^{(2)}(x)$ for all $x\in\ere^d$.
\end{theorem}

 Indeed, it follows from Corollary~4.3  that $V^{(1)}(x)=V^{(2)}(x)$ up to a function from the Schwarz class.
 By our a priori assumption this function has compact support. Then it
 follows from Theorem~4.5  that, actually, this function is zero.

\end{document}